# Influence of Orientational Disorder in the Adsorbent on the Structure and Dynamics of the Adsorbate: MD Simulations of SO$_2$ in ZSM-22

I. Dhiman[a,] *, Sadique Vellamarthodika[b] and Siddharth Gautam[c]

[a]Centre for Energy Research, Budapest, 1121 Hungary

[b] Positron Foundation for Science and Innovation, CITTIC, Cochin University of Science and Technology, Cochin, India

[b]School of Earth Sciences, The Ohio State University, 275 Mendenhall Laboratory, 125 S Oval Drive, Columbus, OH 43210, United States

*Email: indu.dhiman@ek-cer.hu, id.indudhiman@gmail.com

**Abstract**

Structural and dynamical behavior of SO2 molecules within ZSM-22 is studied using MD simulations, to understand the influence of orientational disorder (OD) and inter – crystalline spacing in ZSM-22 as a function of adsorbate loading. Addition of inter-crystalline space provides connectivity of isolated pores in ZSM-22 and is shown to suppress both translational and rotational motion of SO2. We infer that geometry and dimensionality of the connecting space is an important factor in determining the effects of pore connectivity on the adsorbed species behavior. As a function of OD, decrease in self-diffusion coefficient of SO2 in ZSM-22 is observed. An increase in rotational correlation time t and a decrease in libration angle with OD is observed, due to the restriction imposed on the orientational freedom of the adsorbate by an increase in OD. The behavior of SO2 result from an interplay of guest-host interactions and the dimensionality and confinement geometry.

## 1. Introduction

Burning of fossil fuels, several industries and also natural processes lead to emission of toxic gases, such as $CO_2$, $SO_2$, $NO_2$ etc. into the atmosphere [1-9]. The emission of these flue gases from various combustion processes poses serious environmental and health concerns/risks [10-11]. These gases may also undergo several chemical reactions, in turn producing more toxic species and causing further detrimental environmental effects. The concentration of these gases in the environment has increased massively over the last few decades. This has drawn considerable scientific interest towards the development of various techniques, for the separation and capturing / absorption from the emitted flue gases.

Compared to other gases, not many studies have been reported towards the understanding of $SO_2$ separation from flue gas [12-19]. Despite $SO_2$ being one of the most ubiquitous components of fuel combustion, its containment still remains a challenging issue [20-25]. In addition to the toxicity related issues of $SO_2$, its presence in the environment has negative effects in $CO_2$ capturing processes [26-30], one of the main contributors to climate change and global warming [31-37]. To contain these gases nano-porous absorbent materials are observed to have high potential, commonly used for gas separation and catalysis processes. Several absorbents have been studied and reported in literature, for example, nanotubes [38-42], metal organic framework [43-56], zeolite systems [57-63] etc. Among these, zeolites are one of the promising absorbent materials for such applications with several desirable properties, such as large surface area, thermal stability [64] etc. ZSM-22, a high silica zeolite, has a structure that contains sub-nanometer channel-like pores aligned parallel to each other [65]. These nano-porous absorbent materials naturally occur in powder/polycrystalline granular form with grains separated by the inter-particle spacing. The grains are made up of several small crystallites with inter-crystalline spacing, and are randomly oriented with respect to each other [66-71]. This leads to a presence of orientational disorder (OD) in the system. For ZSM-22 a simple geometry of individual pore channels and their network provides a good model system to systematically study the effects of OD on the behaviour of confined fluid. Also, most of the experimental work in this field is carried out on polycrystalline samples, whereas the simulated work is often performed on ideal single crystal samples. As a result the role of inter-crystalline spacing and/or orientational disorder is generally ignored. Zeolites (of interest in current work) in particular have often been modelled as perfect crystals. To address this difference, few studies have been reported on non-idealized models of zeolites. For example, in cation – exchange based models the crystalline imperfections are introduced with the systematic variation of cation content of Zeolites with finite Si/Al ratio [72, 73]. A recent study by Vellamarthodika and Gautam showed the importance of intercrystalline spacing and OD in the absorption of $SO_2$ in ZSM-22 using grand canonical Monte Carlo (GCMC) simulations [74]. The results showed an enhancement in absorption capacity as a function of intercrystalline spacing and OD. While the existence of inter-crystalline spacing and OD have been found to affect the adsorption properties, their effects on the structure and dynamics of the confined species can be expected to be stronger and multifaceted as they provide hindrance to the motion of adsorbed species in the form of discontinuities in the channel-like pore network. To address this, we have investigated the role of intercrystalline spacing in combination with OD in ZSM-22 system on the behavior of confined $SO_2$.

Based on the environmental concerns related to $SO_2$ and its influence on the other gases present in the environment [19-21], we have performed molecular dynamics (MD) simulations to study the structure and dynamical behavior of $SO_2$ in ZSM-22. These simulations are carried out in the presence of inter-crystalline spacing within the crystallites as a function of loading (no. of molecules in a simulation cell) and different extents of orientation disorders. Details related are described in the next section. Our aim is to understand and address these questions in particular, the role of OD in the diffusivity of $SO_2$ within the ZSM-22 pores and the influence of different amount of $SO_2$ adsorbed correlated with the dynamics in ZSM-22. The results obtained from MD simulations make an important scientific contribution towards the understanding of $SO_2$ dynamics in realistic samples (powder like vis-a-vis perfect crystalline), essential towards the understanding of $SO_2$ storage capacity of ZSM-22 adsorbent.

## 2. Simulation Details

The general-purpose software DL_POLY_5.1.0. is used to perform classical molecular dynamics simulations [75]. The ZSM-22 crystal structure is build using the atomic coordinates given by Kokotailo et al. [65] First to construct a single crystallite, the unit cell is replicated 2 × 1 × 3 times with the help of visualization software VESTA [76]. Thereafter, each crystallite was further replicated 2 × 2 × 2 times with an inter-crystalline space inserted between any two neighboring crystallites, forming a complete ZSM-22 simulation cell with eight crystallites separated from each other by inter-crystalline space of width 4 Å. Periodic boundary conditions are applied in all directions. The constructed supercell therefore constitutes a system with infinitely repeated crystallites, with uniform and fixed inter-crystalline spacing along all three directions. For comparison, a simulation cell without any inter-crystalline space between the eight crystallites is prepared and represents an ideal ZSM-22 model. We name the models with and without the inter-crystalline space as IG (Inter-crystalline Gap) and NG (No Gap), respectively. Further, to investigate the influence of orientational disorder in IG ZSM-22 model, individual crystallites are selectively rotated by 90°. To vary the degree of orientation disorder, initially for lowest OD only one crystallite (out of eight in total) is rotated (OD_1), similarly for OD_2, OD_3 and OD_4, 2, 3, and 4 crystallites are rotated, respectively. These models are respectively named IG-OD_1, IG-OD_2, IG-OD_3, IG_OD_4 and represent progressively higher degrees of orientational disorder. A schematic of the resulting simulation cell is shown in Figure 1. The $SO_2$ molecules are added at loadings of n = 16, 32, 48, and 64 molecules in the entire simulation cell with eight crystallites. Both ZSM-22 and $SO_2$ are treated as rigid bodies.

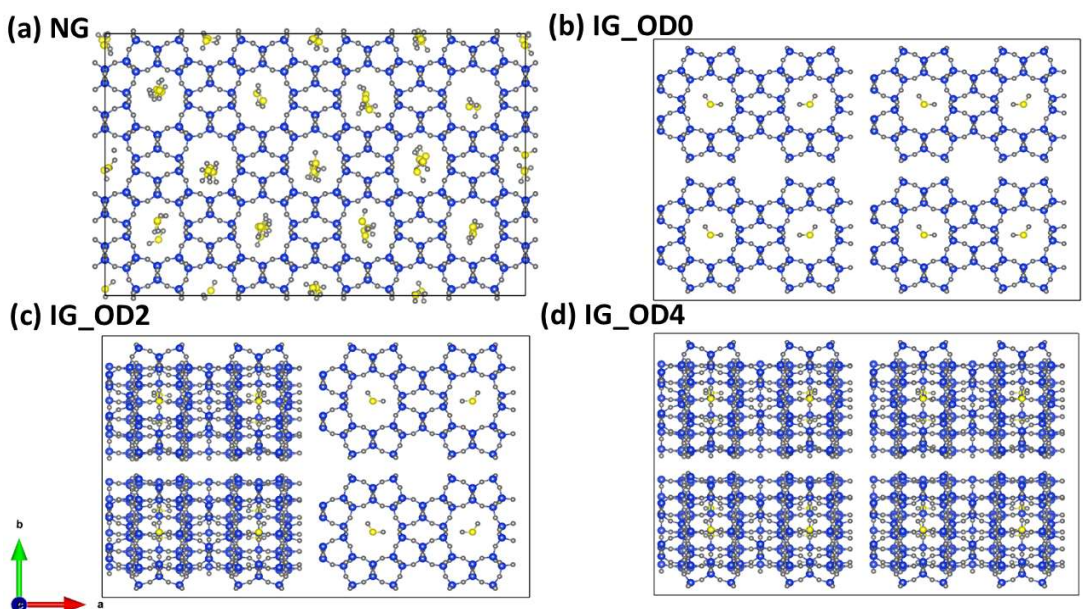

**Figure 1:** (a) Simulation unit cell without inter-crystalline spacing i. e. no gap (NG) and orientational disorder (OD) in the X-Y plane. The black outline shown in the figure marks the unit cell boundary. (b) Simulation unit cell with inter-crystalline gap (IG) i. e. IG_OD0, depicts the inter-crystalline spacing of 4 Å inserted around all the six surfaces. No OD is present in this unit cell. (c) Simulation unit cell IG_OD2, with 4 Å of inter-crystalline spacing and OD is induced by rotating 2 crystallites by 90°. Note that only 2 crystallites that are on the front side (left) of the simulation cell are rotated while all 4 crystallites at the back side remain un-rotated. (d) Simulation unit cell IG_OD4, with 4 Å of inter-crystalline spacing and OD is induced by 4 rotating crystallites by 90°. All the figures shown here are for loading of 64 molecules. Similarly, for IG_OD1 and IG_OD3 (not shown here) 1 and 3 crystallites are rotated by 90° to induce OD, respectively.

NVT ensemble-based simulations are carried out for 6 ns, with the trajectories saved at an interval of 0.5 ps. An integration step of 1 fs is used for simulations. The initial 1 ns run is considered as the equilibration time and the trajectories during this time are discarded. This duration of equilibration time is estimated based on the range where temperature and energy fluctuations are reduced below 5% (approx.). As a result, 5 ns is taken as a production run. The obtained trajectories are unwrapped with visual molecular dynamics (VMD) software [77]. 3D Ewald sum is utilized to calculate long range interactions. To model $SO_2$ we use the parameters employed by Ribeiro [20]. Sun and Han [78] also used these parameters to simulate $SO_2$ in MFI and 4 Å zeolites. $SO_2$ molecules in this formalism is represented by a Sulphur (S) atom rigidly bonded to two Oxygen (O) atoms with a fixed bond length of 1.4321 Å each. O–S–O angle is fixed at 119.5°. To model the interactions of $SO_2$ with ZSM-22 we used the CLAYFF force-field parameters [79] to represent Silicon and Oxygen atoms. In our initial configuration for each crystallite unit cell equal amounts of molecules are placed in the center of each pore, as evident in the Figure 1.

## 3. Results

3.1 Structural Behavior

To understand the structural behavior, we have plotted the intensity maps showing the distribution of Sulphur belonging to $SO_2$ molecule in ZSM-22 frame. Figure 2 shows the corresponding intensity maps in XY plane for loading of 64 molecules for all the systems. In Figure 2 the Z-intensity is log ($N$ + 1), where $N$ is the number of Sulphur atoms of $SO_2$ molecule occupying a particular location (projection in the X–Y plane) at some instant during production run. The intensity distribution here gives the relative probability of finding a Sulphur atom at a given location. In the top rows of Figure 2 Sulphur distribution in the entire simulation cell is displayed, while the lower row of the same figure shows the close - up view of one representative pore.

For the simulation cell without inter-crystalline spacing and OD i. e. in NG (Figure 2(a)) the $SO_2$ molecules prefer to occupy the center of the ZSM-22 pore. With the introduction of inter-crystalline spacing of 4 Å and in the absence of OD for IG_OD0, the $SO_2$ molecules show slight redistribution. Few molecules reside in the introduced inter-crystalline spacing in addition to occupying the center of the ZSM-22 pore. This redistribution of $SO_2$ molecules is more clearly visible in the intensity distribution plotted in YZ - projection/plane, as shown in the supplemental Figure S1. In the next simulated system (IG_OD1) with the introduction of OD by rotating one crystallite out of eight accompanied with inter-crystalline spacing, $SO_2$ molecules present around the OD surroundings begin to delocalize and also occupy the periphery of the inter-crystalline gap. With increase in OD for IG_OD2, with two out of eight crystallites rotated, further $SO_2$ molecules are delocalized and prefer to occupy the inter-crystalline gaps. Further increase in OD for IG_OD3 and IG_OD4 the intensity of $SO_2$ molecules within the inter-crystalline gaps is also increased, implying further delocalization of molecules induced by higher OD present in the system.

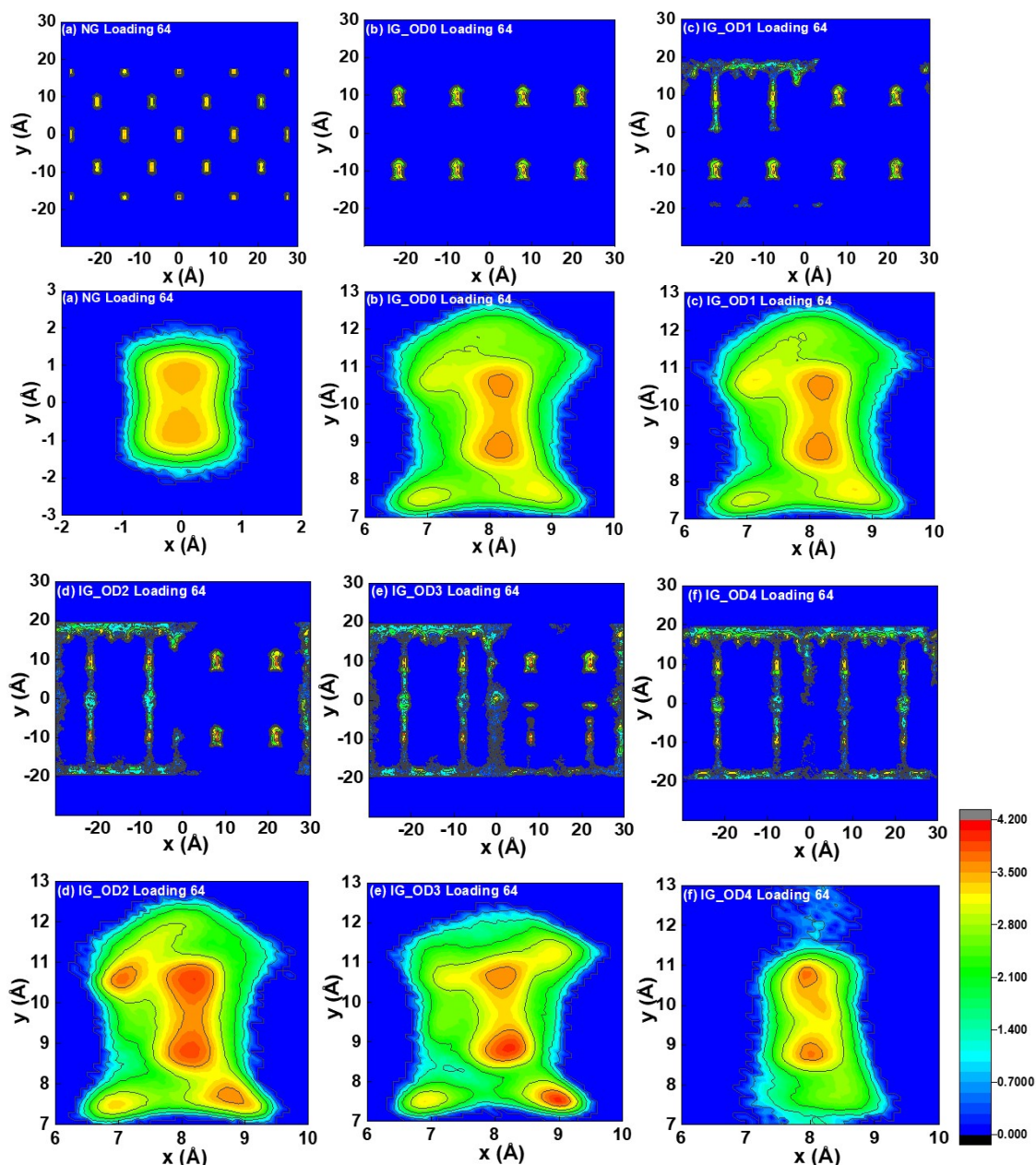

**Figure** 2: Probability distribution of Sulphur atoms belonging to SO$_2$ molecules in the X – Y plane, with loading n = 64 molecules for (a) NG, (b) IG_OD0, (c) IG_OD1, (d) IG_OD2, (e) IG_OD3 and (f) IG_OD4 in ZSM-22. The I and III row of figures show the SO$_2$ distribution within the whole simulated unit cell, while in II and IV row figures are the zoomed in view of one representative pore. The Z-intensity color code shows the value log (N + 1), where N corresponds to the number of SO$_2$ molecules (i. e. Sulphur atoms) occupying a particular location for some instant during the whole production run time.

## 3.2 Orientational Distribution

Orientational structure of the adsorbed $SO_2$ is studied with reference to the angle (θ) made by the molecular axis (S-O vector) with Cartesian directions. Distribution of molecular populations over different angles ranging between 0 and 180 degrees is calculated and termed as orientation distribution function (ODF). Figure 3 depicts the ODF for (a) NG, (b) IG_OD0, (c) IG_OD1, (d) IG_OD2, (e) IG_OD3 and (f) IG_OD4 in ZSM-22 along x-, y- and z- axis at loading of 64 molecules. Also included in the plots is the ODF expected for an isotropic system with no preferred direction. Deviation of the ODF from the isotropic case shows the extent of preference shown by $SO_2$ molecules towards a particular orientation, or orientational anisotropy. Similar behavior is observed for all the systems as a function of their respective loadings. In particular, for NG system in Figure 3(a), along X- axis a sharp peak at 90° is observed, indicating the molecule is aligned perpendicular to X- axis. While, along Y- axis two symmetric broad peaks around 90° are observed, implying preferred orientation. In contrast, along z – axis the ODF nearly overlaps with the isotropic one implying no preference in orientation with respect to the pore axis. With the introduction of inter-crystalline spacing and OD, the $SO_2$ molecules exhibit signature of preferred orientation along all x-, y- and z- directions. For IG_OD0 in Figure 3(b), with the introduced inter-crystalline spacing, the ODF is no longer symmetric in nature around 90°, in contrast to the behavior observed for NG system (Figure 3(a)). As a function of increasing OD for IG_OD1, IG_OD2, IG_OD3 and IG_OD4 simulation cells, the asymmetry in ODF is weakly but consistently increased. This consistent increase in the orientational anisotropy has important consequences for the rotational motion as we shall see.

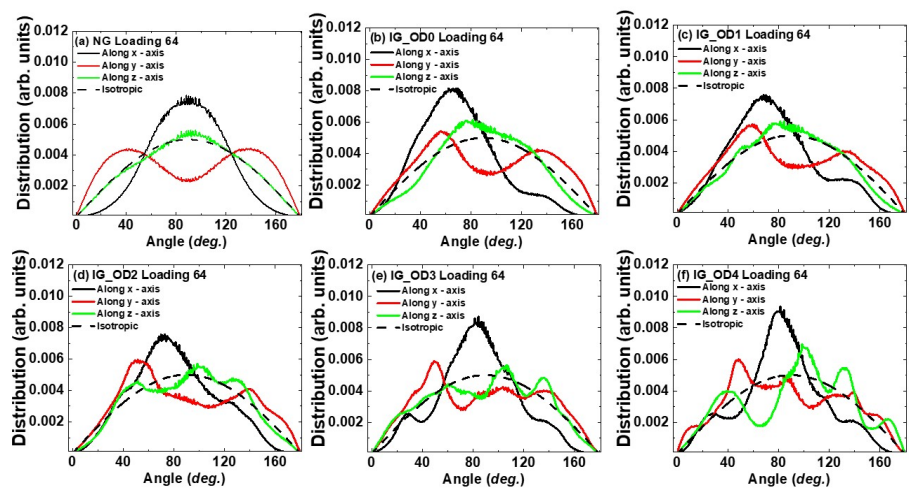

**Figure 3**: Orientation distribution of the S-O vector of $SO_2$ molecules (a) NG, (b) IG_OD0, (c) IG_OD1, (d) IG_OD2, (e) IG_OD3 and (f) IG_OD4 in ZSM-22 with respect to X- (black solid line), Y- (red solid line) and Z- (blue solid line) axis for n = 64 loading is shown. The black dashed lines shown in all the figures provides comparison of the expected orientational distribution for an ideal isotropic behavior.

## 3.3 Translation Dynamics

The translation dynamics of SO$_2$ molecules in ZSM-22 is understood by calculating the mean square displacement (MSD) of Sulphur atoms as a function of time. Note that for a small rigid molecule like SO$_2$ the evolution of this MSD is close to that of the MSD of the molecular center of mass (COM). The time dependence of MSD behavior for all the loadings and NG, IG_OD0, IG_OD1, IG_OD2, IG_OD3 and IG_OD4 systems is shown in the supplemental figure S2. Linear dependence of MSD as a function of time is observed for all the systems. For NG, IG_OD0, IG_OD1, and IG_OD2 systems, reduction in MSD with increasing loading is observed, while for IG_OD3 all MSDs nearly coincide with each other for all loadings. Interestingly, for IG_OD4 system a trend with MSD displaying a maximum for loading 32 molecules as compared to the Loadings of 16, 48 and 64 molecules is observed. This overall behavior of MSD shows that as a function of increasing orientational disorder MSD is reduced.

To understand the directional dependence related to the mobility of SO$_2$ molecules the MSD components with respect to the Cartesian x-, y- and z- directions are calculated. The pertaining figures are shown in the supplemental Figure S3. Initially for NG and lower OD systems i. e. for IG_OD0, IG_OD1, and IG_OD2 the MSD along z- axis exhibits higher values, in comparison to x– and y– directions. However, the difference in MSD between z- axis and x-, y- axis is narrowed with increasing OD. For systems with higher disorder, for IG_OD3 and IG_OD4, the difference between MSDs along all the Cartesian directions is significantly narrowed. For NG system, in the absence of an inter-crystalline gap, the movement of SO$_2$ molecules is facilitated only along the z- direction as all the channel-like pores are oriented along this direction. With the introduction of inter-crystalline spacing and absence of OD for IG_OD0, SO$_2$ molecules become comparatively free to move along x - and y- axis in addition to the z- axis. This leads to an increase in MSD values along x- and y- axis, therefore narrowing the difference between z- axis and x-, y- axis. Thereafter, with the introduction of OD in IG_OD2 and increase in its extent for IG_OD3 and IG_OD4, progressively more channels running perpendicular to the Cartesian Z-direction become available for molecules to move, therefore the difference in directional MSDs is further reduced. Two effects occurring here lead to an overall (also relative to z- axis) enhancement in SO$_2$ mobility along x- and y- directions: (1) insertion of inter-crystalline gap and (2) the presence of orientational disorder leads to a redistribution of MSD contribution along z - direction to a perpendicular plane because some pores that were originally in this direction have now been turned perpendicular to it.

Self-diffusion coefficient (D) as a function of loading for overall system and along x-, y- and z- directions is depicted in Figure 4. The diffusion coefficient is calculated by fitting the Einstein equation (eqn. 1 below) to MSD for the time scales above 100 ps. This time range is much above the initial ballistic region, i. e. the initial range of few picoseconds.

The Einstein equation is given as

$$D = \frac{MSD}{2n_d t} \quad \text{------------- (1)}$$

where diffusion coefficient is given as D and $n_d$ represents the dimensions of space in which the motion occurs.

The diffusion coefficient as a function of loading for overall MSD is shown in Figure 4(a). For clarity the y- axis is shown in log scale. For all the simulated systems, NG, IG_OD0, IG_OD1, IG_OD2, IG_OD3 and IG_OD4, as a function of loading overall diffusion coefficient shows a decrease. This behavior may be attributed to the crowding of the molecules with increase in number of $SO_2$ molecules (i. e. loading), therefore restricting the motion. Likewise, with increasing OD diffusion coefficient exhibits a reduction, which may be ascribed to the hindrance caused by the induced orientational disorder. In Figure 4(b), (c) and (d) the diffusion coefficient variation with loading is shown along x-, y- and z- directions, respectively. Similar to overall diffusion coefficient behavior, reduction in D values is observed along all the directions for all the simulated unit cell systems. Further, it is noticeable that the overall diffusion coefficient value is dominated by the $SO_2$ motion along z- direction.

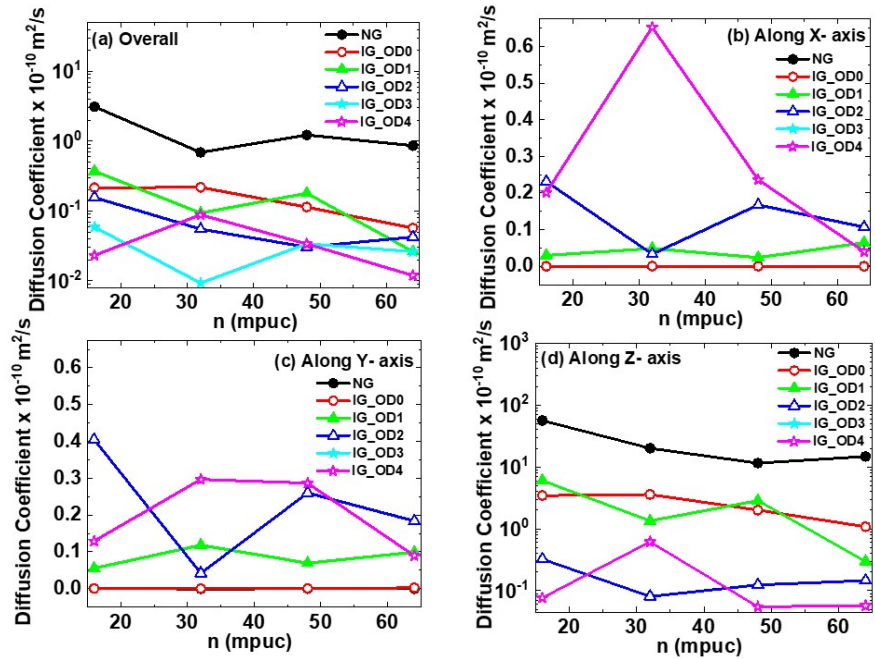

**Figure 4:** Direction dependent diffusion coefficient as a function of loading for (a) overall, (b) x, (c) y-, and (d) z- directions for NG, IG_OD0, IG_OD1, IG_OD2, IG_OD3, and IG_OD4 simulated systems. For clarity diffusion coefficient for overall and along z- direction is shown in log scale in y-axis.

3.4 Rotational Dynamics

To understand the rotational dynamics of $SO_2$ molecules, the rotational correlation function (RCF), $C_R(t)$, is calculated using the equation,

$$C_R(t) = \langle \mathbf{u}(t + t_0) \cdot \mathbf{u}(t_0) \rangle \text{ -------------------- (2)}$$

where, $\mathbf{u}$ is the unit vector defined with respect to the molecular axis, the $\langle \rangle$ brackets shows that the $C_R(t)$ is calculated by taking average over all the molecules and time origins ($t_0$). The RCF function can be compared with experimentally obtained data using dielectric / IR-spectroscopy. Figure 5(a) displays the time dependent variation of $C_R(t)$ for all the simulated unit cells with loading 64 molecules. All the RCFs exhibit an initial fast decay in the sub-picosecond region followed by a much slower decay at longer times. The two regimes of the fast and slow decay are marked by a boundary at around 1 ps, where wiggling of the RCF can be observed for all systems. This is a signature of short time fast librational motion. It is noteworthy that the value of RCF at the boundary of the two regimes is significantly different for different systems. The location of local minimum in RCF at the boundary is a measure of the angle of librations [80]. The RCF for different systems suggest that the angle of libration is highest for NG and gets progressively reduced with the introduction of inter-crystalline space and OD. In the long-time regime, for NG, RCF behavior shows relatively fast decay, with complete decay occurring within few tens of picosecond. This behavior indicates the fastest rotational dynamics observed of $SO_2$ molecules in the absence of inter-crystalline spacing and OD. When inter-crystalline spacing is introduced in IG_OD0, RCF decay is slightly suppressed, indicating the slowing down of molecular rotation. The decay rate is progressively reduced for higher OD systems (from IG_OD1 to IG_OD4), indicating the restrictive nature of molecular rotation induced by an increase in the extent of ODs. The time dependence of RCF for all the loadings is shown in the supplemental Figure S4. To further elaborate, we have obtained the time scales of rotational motion by fitting the RCF above 1 ps with an exponential decay function, with the rotational correlation time ($\tau$) as the fitting parameter. The rotational correlation times obtained thus are shown in Figure 5(b). For NG and IG_OD0 reduction in $\tau$ as a function of loading is seen. For systems with OD i. e. IG_OD1, IG_OD2, IG_OD3, and IG_OD4, no significant variation in $\tau$ with loading is observed. However, for each value of loading $\tau$ exhibits nearly continuous increase with orientation disorder.

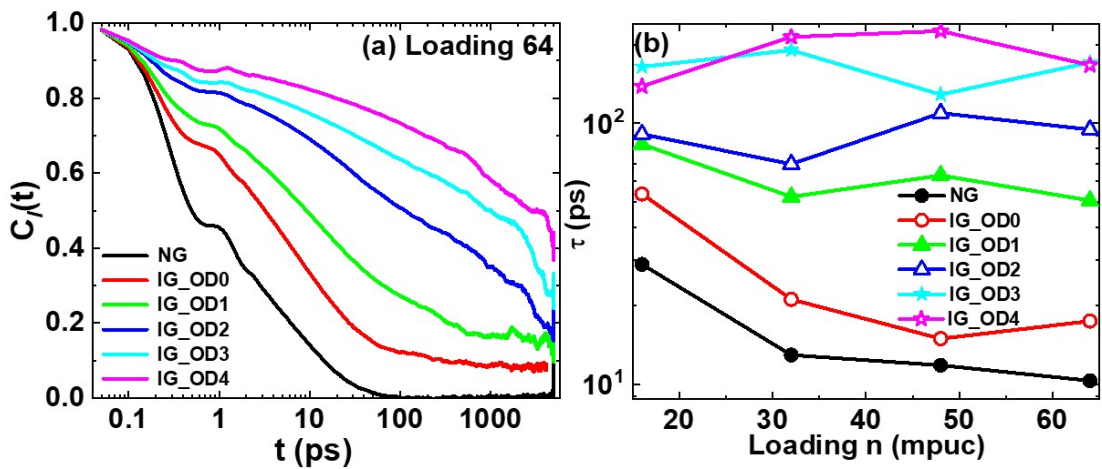

**Figure 5:** (a) The rotational correlation as a function of time for Loading 64 and (b) the loading dependence of orientational correlation times obtained from rotational correlation functions for NG, IG_OD0, IG_OD1, IG_OD2, IG_OD3, and IG_OD4 simulated unit cells.

## 4. Discussion

ZSM-22 with its nm range porous structure is a promising absorbent for $SO_2$ adsorption. Considerable influence of inter-crystalline spacing and orientational disorder is ascertained in the adsorption of $SO_2$ in ZSM-22. Particularly, the preferred adsorption on the surface of ZSM-22 crystallites is enhanced with the introduction of inter-crystalline spacing and OD. In the probability distribution plot shown in Figure 2(a) for NG system with loading 64, $SO_2$ molecules distribution follows the shape of ZSM-22 pores along z - axis, i. e. $SO_2$ molecules are highly localized. With the introduction of inter-crystalline spacing in Figure 2(b) for IG_OD0 with loading 64 system, the molecules also begin to redistribute along x – and y – axis outside the pore boundaries. This redistribution continues with the introduction of OD and with variation of its extent from IG_OD1 to IG_OD4 (Figure 2(c-f)). The influence of OD is also clearly seen in ODF (Figure 3), MSD (Figure 4), and RCF (Figure 4) behaviors.

Model ZSM-22 in its crystalline form (NG here) has channel-like pores that run along the Z-direction infinitely while being isolated from each other with no opportunity for an adsorbed molecule to migrate from one pore to the other. When inter-crystalline space is inserted in the model IG_OD0, these pores get connected resulting in the possibility of inter-pore migration of the adsorbed molecules. While to our knowledge this is the first study addressing the effects of orientational disorder, the effects of connecting the pores in ZSM-22 by inserting inter-crystalline space in the X-Y plane, on the behavior of ethane and $CO_2$ have been reported [81]. While the effect of the inter-crystalline space added thus was a suppression of rotational motion of both ethane and $CO_2$, the translational motion of the two adsorbates was affected differently. Ethane molecules were found to exhibit a faster translation motion while $CO_2$ mobility was lowered on connecting the pores via this inter-crystalline space. The behavior of $SO_2$ reported here is thus similar to the behavior of $CO_2$ reported by Kummali et al. [81]. This could be due to the strong electrostatic interactions between the guest ($SO_2$ or $CO_2$) and the host which are absent in the case of ethane.

To illustrate the changes in $SO_2$ - host interactions with respect to the inter-crystalline spacing and orientational disorder, the overall behavior is pictorially summarized in Figure 6. For both NG and IG_OD0 systems all the cylindrical pores of ZSM-22 are aligned along z – axis, as shown schematically in Figure 6(i) for IG_OD0. When orientational disorder is induced in the system, some of these cylindrical pores are reoriented, a schematic is shown in Figure 6(ii). As a consequence of this, few of these cylindrical pores of ZSM-22 are no longer aligned along z – axis. This obstructs the uninterrupted cylindrical path initially available to $SO_2$ molecules movement. For IG_OD1 only fewest of these cylindrical pores are reoriented, i. e. OD is smallest for IG_OD1. Therefore, diffusion coefficient shows only slight reduction along z – axis (Figure 4(d)). In contrast, as a function of increasing OD in IG_OD2, IG_OD3, and IG_OD4 systems, more of the crystallites are progressively reoriented. Such that more of these cylindrical pores are obstructed for the $SO_2$ movement. The anisotropic nature of ODF is consistently increased and increase in ODF as a function of OD occurs. The rotated crystallites as a result of OD restricts the orientational space available for $SO_2$ molecules. This may imply that in the presence of OD the rotation motion of $SO_2$ molecules become restrictive (with smaller orientational space available), with the extent of restrictiveness increasing as a function of OD.

Diffusion coefficient for different systems for loading 64 is shown in Figure 6(iii). As expected, the presence of OD clearly affects the translation motion of $SO_2$ molecules, with diffusion coefficient reducing with increasing OD. We note that for some other loadings (not shown in Figure 6(iii)) some deviation from monotonous behavior is observed. However, in general, translation motion is observed to slow down at higher OD. Also, the value of diffusion coefficient is dominated by the values along z –axis. More interestingly, rotational dynamics exhibits a strong dependence on the OD. $\tau$ as a function of OD (different systems shown on X-axis represent OD increasing to the right) for loading 64 is displayed in the inset to Figure 6(iii). The figure shows an increase in $\tau$ with increase in OD, implying restriction to rotational motion of $SO_2$ molecules at higher OD. To further illustrate the role of OD on the rotational motion of $SO_2$ three scenarios are pictorially depicted in the Figure 6(iv)a, b and c. In the first case shown Figure 6(iv)a the $SO_2$ molecule is present inside the cylindrical pores of ZSM-22. This is the most probable case for NG and IG_OD0 systems. Here, a typical $SO_2$ molecule experiences uniform interaction within the cylindrical pore walls of ZSM-22; therefore, the absence of preferred orientation (nearly isotropic ODF with respect to Z-direction in Figure 3 (a and b)). However, when the OD is introduced in IG_OD1, IG_OD2, IG_OD3, and IG_OD4 systems, molecules begin to translate in the inter-crystalline spacings and also experience obstruction owing to the presence of OD, as shown in Figure 6(iv) b and c. This favors the preferred orientation of $SO_2$ molecules as seen in Figure 3 (c – f).

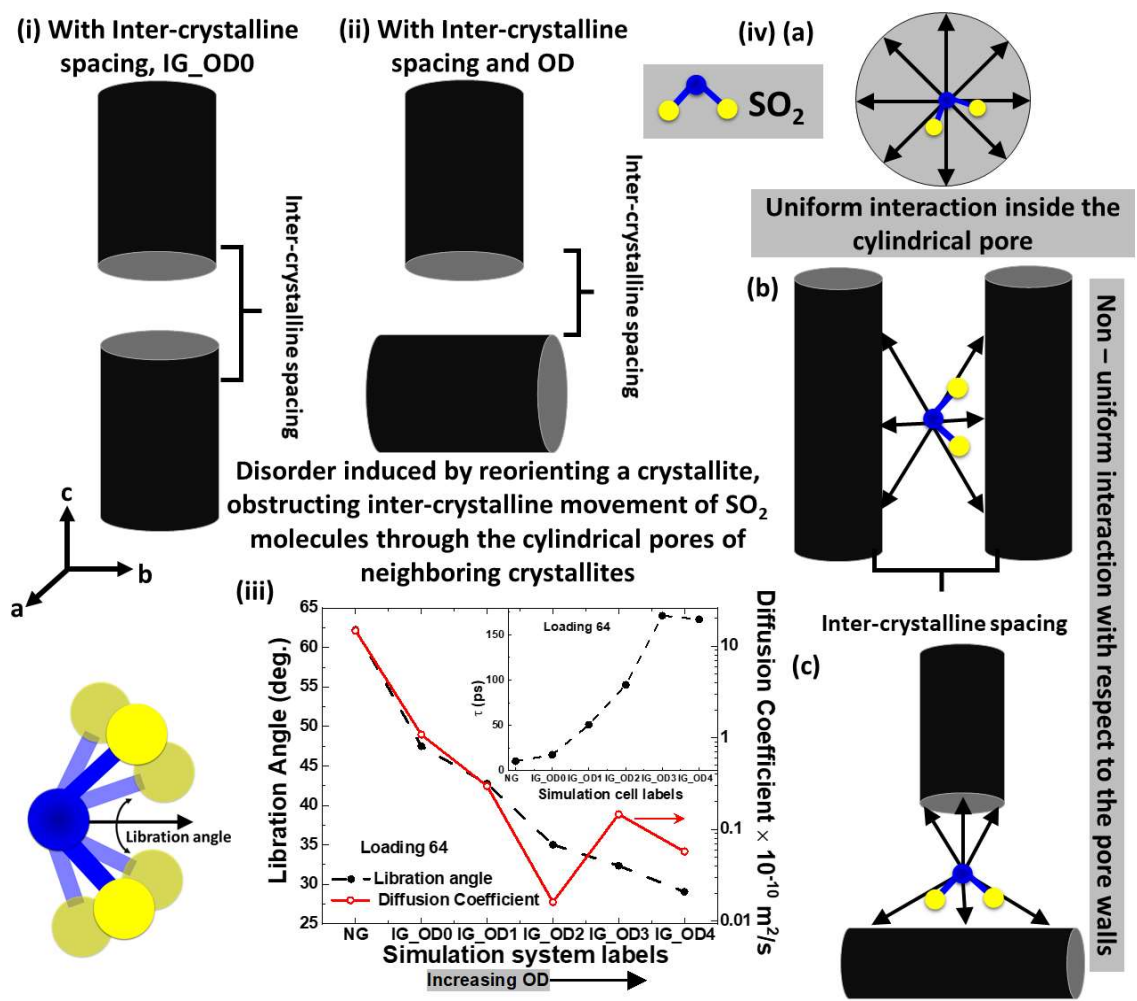

**Figure 6:** A pictorial representation of $SO_2$ absorption in ZSM-22. (i) In the absence of orientational disorder (OD) and with inter-crystalline spacing, the black cylindrical shape represents the pores in ZSM-22. (ii) When in addition to inter-crystalline spacing OD is introduced, the pores are realigned as shown. (iii) Libration angle and diffusion coefficient for different systems for loading 64, while the inset to the figure shows the rotational correlation time $\tau$ for loading 64. The systems on the X-axis represent progressively increasing OD with the two leftmost systems representing an absence of OD. (iv) Interaction experienced by the $SO_2$ molecules when present (a) within the cylindrical pores, (b) in the inter-crystalline space of two parallelly aligned pores and, (c) in the inter-crystalline space when one of the ZSM-22 pore is reoriented at 90° w. r. to the neighboring pore.

We also calculated the libration angle of SO$_2$ molecules shown in Figure 6(iii) for loading 64. The libration angle is calculated using the first order OCF, shown in Figure 6(a). To obtain the libration angle, the inverse cosine of C$_l$(t) value is taken at the first minimum [82] observed in Figure 6(a). For NG system, SO$_2$ molecules mostly reside inside the cylindrical pores and show the highest libration angle value. This can be correlated with the maximum freedom the SO$_2$ molecules have within these cylindrical pores without any preferred orientation, also pictorially depicted in Figure 6(iv) a. On the other hand, in the presence of inter-crystalline space and OD, the libration angle exhibits reduction as a function of OD (Figure 6(iii)). This implies the hindrance in the libration of the molecules and also preferred orientation induced as a result of OD.

The inter-crystalline space is of similar width as the pore. The pores with cylindrical shape impose confinement in two directions (i. e. along x – and y - direction) and can therefore be said to offer 2-D confinement, while the inter-crystalline space with somewhat slit like pores imposes confinement only in one direction, or in other words, imposes 1-D confinement. In principle this implies that geometrically more space is available within the inter-crystalline space, as against the pores. This difference in the geometry and dimensionality of the pores and the inter-crystalline space has important consequences for the dynamics in general. As stated above, inserting the inter-crystalline space between ZSM-22 crystallites is a means of connecting the isolated channel-like pores. This connectivity via 1-D confining inter-crystalline space is found to suppress both the translational as well as rotational motions of SO$_2$ reported here and CO$_2$ earlier [83]. In another study on the effects of pore connectivity where cylindrical pores imposing confinements in two directions were connected by similar 2-D confining pores, the translational as well as rotational motions of both ethane and CO$_2$ were enhanced with pore connectivity [84]. The effects of pore connectivity are thus dependent on the geometry and dimensionality of the connecting space.

Moving from 2-D confining cylindrical pores to 1-D confining inter-crystalline space, one would expect a relaxation in libration angle, contrary to what we observe in Figure 6(iii). The observed reduction in the range of libration angle occurs because of the stronger interaction of SO$_2$ with the crystallite surface, also in agreement with the GCMC study reported earlier [72]. This shows the importance of the dimensionality and geometry of confinement and the role of guest-host interactions in determining the behavior of the confined species. In general, it can be surmised that the effects of confinement on the behavior of adsorbed species may depend on the dimensionality of confinement and hence confinement in a slit pore might have effects different from confinement in cylindrical pores.

Sabahi et al. reported a molecular dynamic simulation study of SO$_2$ in silica Y zeolite [84]. The silica Y zeolite and ZSM-22 are of same chemical composition, albeit of different pore shape and dimension. The pores in the silica Y zeolite are larger. and essentially of spherical shape connected by narrow windows. Interestingly, the diffusion coefficient values for SO$_2$ in silica Y zeolite is a nearly order of magnitude larger than the values observed in the current work for ZSM-22. Since the chemical interaction in the two systems are rather similar, the observed difference in the diffusion coefficient values between the two systems can be ascribed to the different geometrical confinement imposed by them on $SO_2$.

## 5. Conclusion

In the current work, we present the results of molecular dynamics simulations investigating the influence of orientational disorder (OD) and the presence of inter – crystalline spacing as a function of loading of $SO_2$ molecules within the ZSM-22. Addition of inter-crystalline space provides connectivity of isolated pores in ZSM-22. Comparing with other studies on the effects of pore connectivity, we infer that the geometry and dimensionality of the connecting space is an important factor in determining the effects of pore connectivity on the behavior of adsorbed species. Significant influence of inter – crystalline spacing and OD on structural and dynamics behavior of $SO_2$ molecules is observed. The introduction of OD and with increase in its extent leads to a reduction in self-diffusion coefficient and an increase in rotational correlation time $\tau$ for $SO_2$ molecules. This behavior implies that the presence of OD hinders both translation and rotational motion of $SO_2$ molecules. To further elaborate, the libration angle is obtained from the time dependent RCF plots. For systems without OD, nearly isotropic orientational distribution of $SO_2$ molecules is observed, i. e. no preferred orientation is found, while, with the introduction of OD, the orientational distribution depicts anisotropic behavior. As a result, a decrease in libration angle as a function of increasing OD is seen. The observations reported here result from an interplay of guest-host interactions and the dimensionality and geometry of confinement. In summary, our study reveals the important role of inter-crystalline spacing and orientational disorder on the structure and dynamics of $SO_2$ molecules within ZSM-22.

## 6. Author Contributions

Conceptualization – I.D. and S.G., Preparing input for MD simulations – I. D., S. V. and S. G., Carrying out MD simulation and data analysis – I. D., Preparing manuscript draft – I. D., Editing manuscript draft – I. D., S. V. and S. G.

## 7. Acknowledgements

We would like to acknowledge STFC's Daresbury Laboratory for providing the package DL-Poly, which was used in this work. Help received in the analysis of MD simulations from the visualization packages VESTA and VMD is also acknowledged. I. D. acknowledges KIFÜ for awarding the access to computational resources based in Hungary at Budapest.

## 8. Conflict of Interest

The authors declare no conflict of interest.

## 9. Funding

This work received no external funding.

## 10. Data Availability

All data supporting the findings reported in this work are included in the article and the associated supplementary information file.

## 12. Supplementary Information

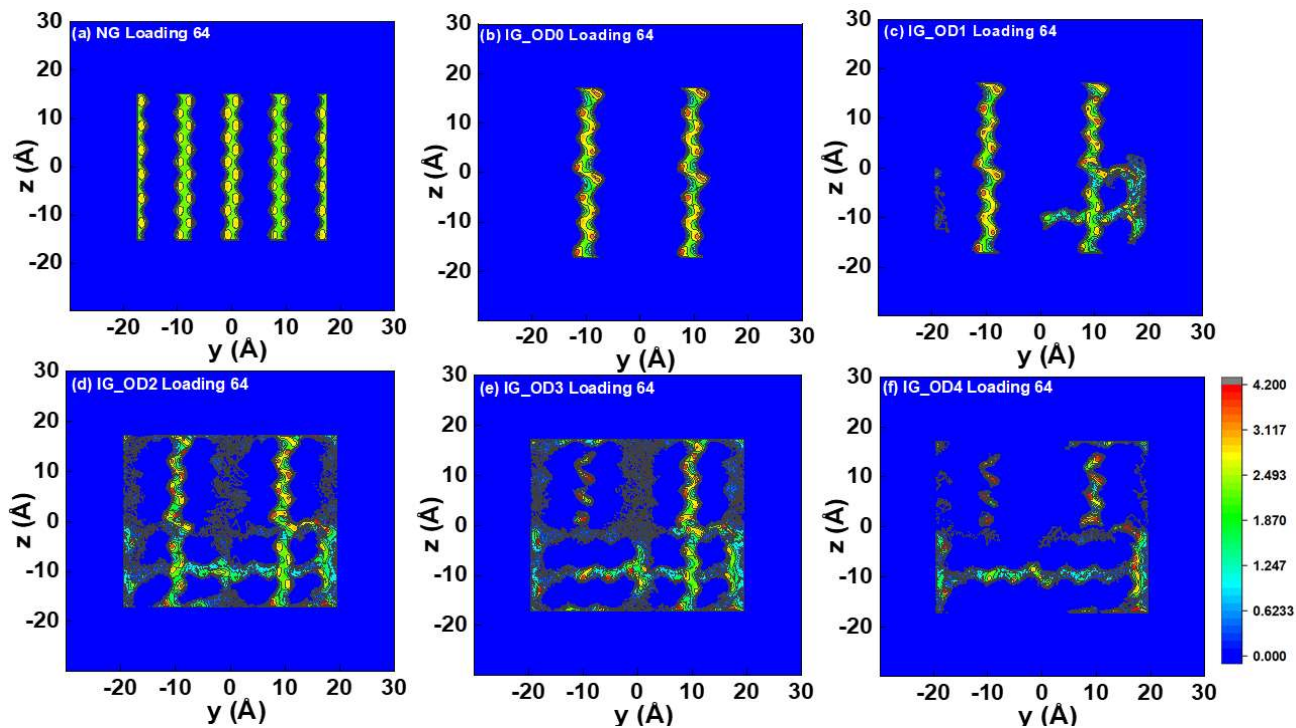

Figure S1: Probability distribution of Sulphur atoms belonging to SO$_2$ molecules in the Y – Z plane, for (a) NG, (b) IG_OD0, (c) IG_OD1, (d) IG_OD2, (e) IG_OD3 and (f) IG_OD4 with loading n = 64 molecules in ZSM-22. The Z-intensity color coding represents log (N + 1), where N corresponds to the number of SO$_2$ molecules (i. e. Sulphur atoms) occupying a particular location for some instant during the whole production run time.

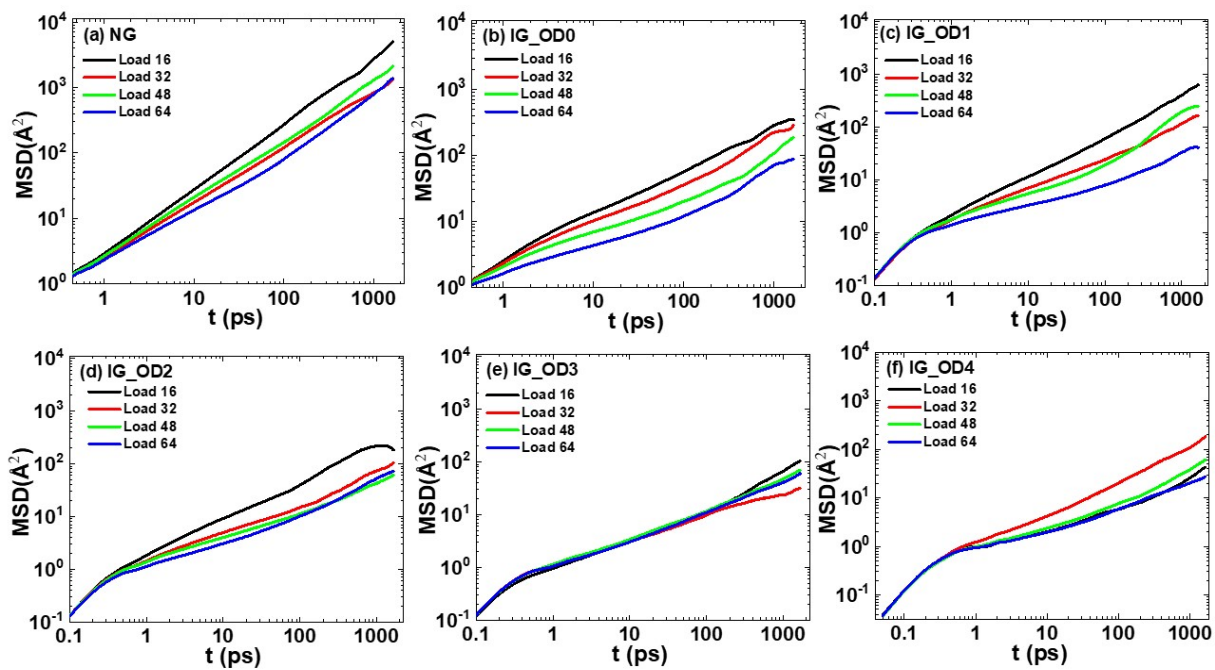

**Figure S2:** Time dependent mean square displacement (MSD) as a function of loadings for (a) NG, (b) IG_OD0, (c) IG_OD1, (d) IG_OD2, (e) IG_OD3 and (f) IG_OD4.

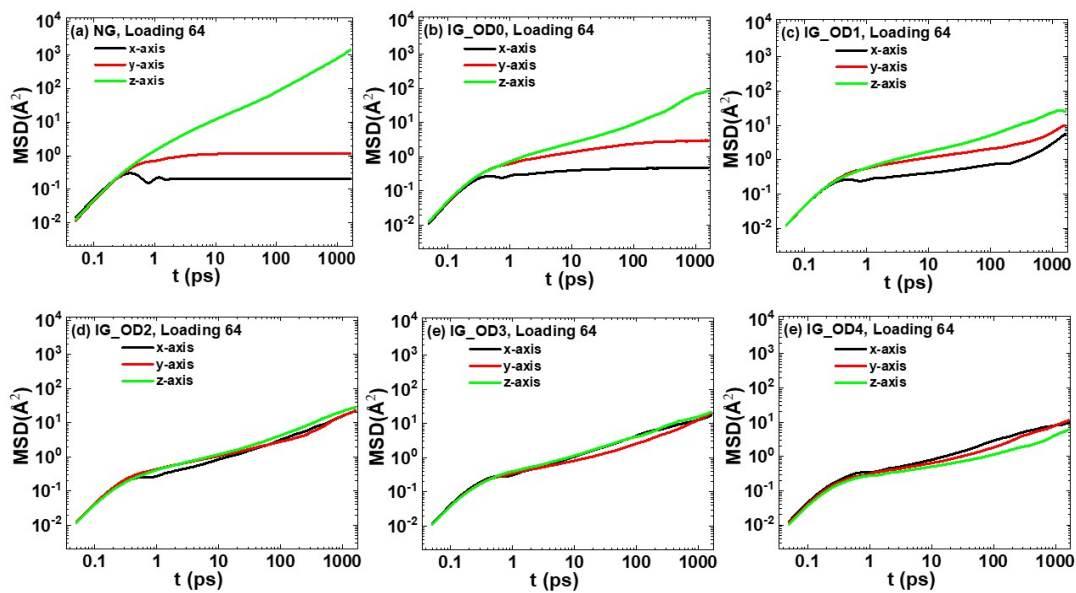

**Figure S3:** Mean square displacement (MSD) as a function of time for loading, n = 64 molecules for (a) NG, (b) IG_OD0, (c) IG_OD1, (d) IG_OD2, (e) IG_OD3 and (f) IG_OD4 along x –, y–, and z– directions. The plots shown here are representative for the respective lower loadings of n = 16, 32, and 48 molecules.

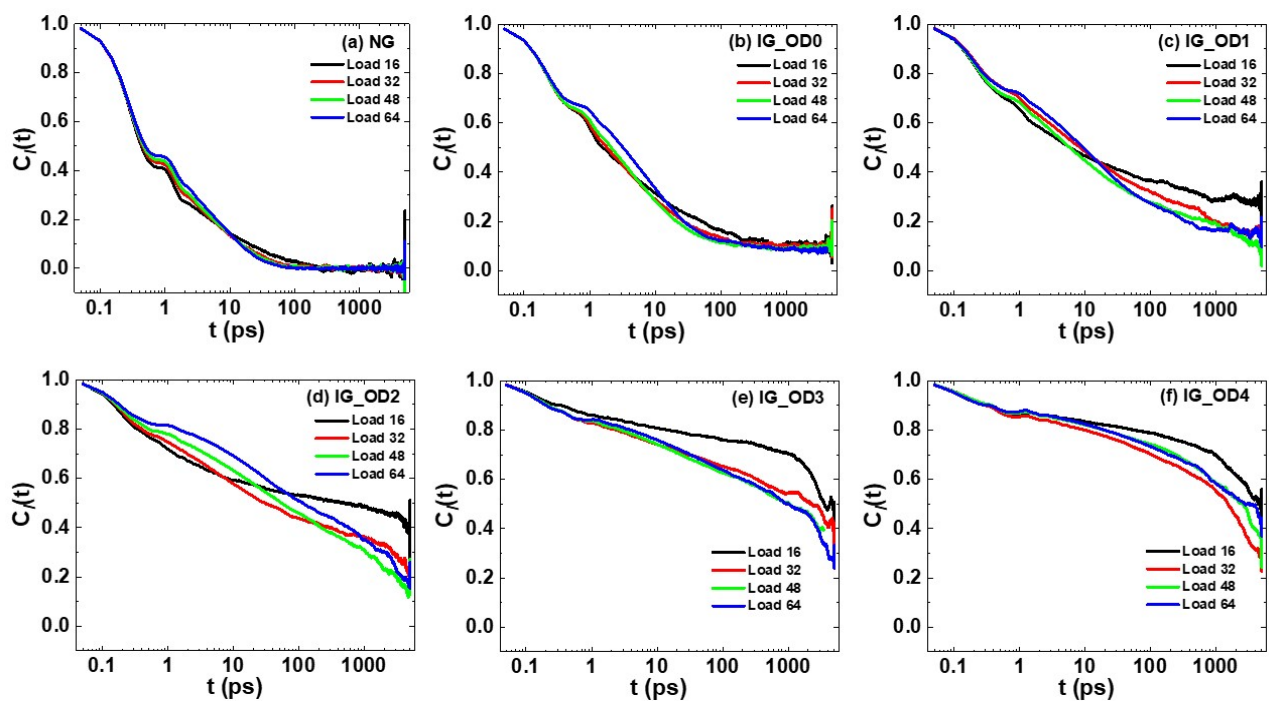

**Figure S4:** The rotational correlation function at different loadings for (a) NG, (b) IG_OD0, (c) IG_OD1, (d) IG_OD2, (e) IG_OD3 and (f) IG_OD4.